# Generation and reverse transformation of twisted light by spatial light modulator

Zhi-Yuan Zhou*,[1,2] Zhi-Han Zhu,[3] Shi-Long Liu,[1,2] Shi-Kai Liu,[1,2] Kai Wang,[1,2] Shuai Shi,[1,2] Wei Zhang,[1,2] Dong-Sheng Ding,[1,2] and Bao-Sen Shi*[1,2]

[1]*CAS Key Laboratory of Quantum Information, USTC, Hefei, Anhui 230026, China*

[2] *Synergetic Innovation Center of Quantum Information & Quantum Physics, University of Science and Technology of China, Hefei, Anhui 230026, China*

[3]*Institute of photonics and optical fiber technology, Harbin University of Science and Technology, Harbin 150080, China.*

*[\*zyzhouphy@ustc.edu.cn](mailto:zyzhouphy@ustc.edu.cn); [drshi@ustc.edu.cn](mailto:drshi@ustc.edu.cn)*

A spatial light modulator (SLM) is one of the most useful and convenient device to generate structural light beams such as twisted light and complexed images used in modern optical science. The unbounded dimension of twisted light makes it promising in harnessing information carrying ability of a single photon, which greatly enhances the channel capacity in optical communications. We perform a detail theoretical study of the birth, evolution and reverse transformation of twisted light generated from a phase-only SLM based on diffraction theory, analytical expressions are obtained to show the special evolution behaviors of the light beam with the propagation distance. Beam intensity distributions calculated theoretically are in well agreement with experimental observations. Our findings clearly reveal how twisted light gradually emerges from a Gaussian spatial shape to a ring shape, and also the ring shape will gradually evolve to a quasi-Gaussian shape conversely, in the reverse transformation. These results will provide guidelines for using SLM in many optical experiments and OAM-mode multiplexing in optical communications.

Recently people have spent great efforts in generating and controlling of structural light beams to use them in modern optical science [1, 2]. Among many structural light beams, a twisted light beam is the most studied for its broad applications including high capacity free space and fiber optical communications[3, 4], micro-particle trapping and manipulation[5], high-precise optical metrology[6-8], study nonlinear interaction processes[9-12] and quantum information science [13-17]. All these applications originate from special properties of twisted light: unbounded dimensions, unique mechanical torch effect and singularity in phase and intensity distributions. A twisted light beam is also called a vortex beam for its spiral phase front of form $e^{il\theta}$, which carries $l\hbar$ orbital angular momentum each photon [18].

Practical applications of vortex beams rely on effective generation and controlling the beam evolution dynamics. The dynamical beam evolution of vortex beams has been broadly studied in both linear [19, 20] and nonlinear [21] transformation processes. Usually, a vortex beam can be generated by a hologram grating, a vortex phase plate and a spatial light modulator (SLM). Among these methods, generating vortex beams by applying computer-generated phase mask to SLM is the most convenient in practical experiments. Though SLM has been widely used in many experiments for generating vortex beams and also being used to reversely transform vortex beams to Gaussian modes to couple them into single mode fibers [14, 16], no one gives a detailed study of beam evolution behaviors in the whole beam transformation process to our best knowledge.

In this article, we give details for vortex beams generation, evolution and reverse transformation of vortex beams to Gaussian modes based on diffractive optical theory. Analytical expressions against propagation of the two processes are obtained, these results give insight on vortex beam generation and propagation, which will be guidelines for researches working on vortex beam.

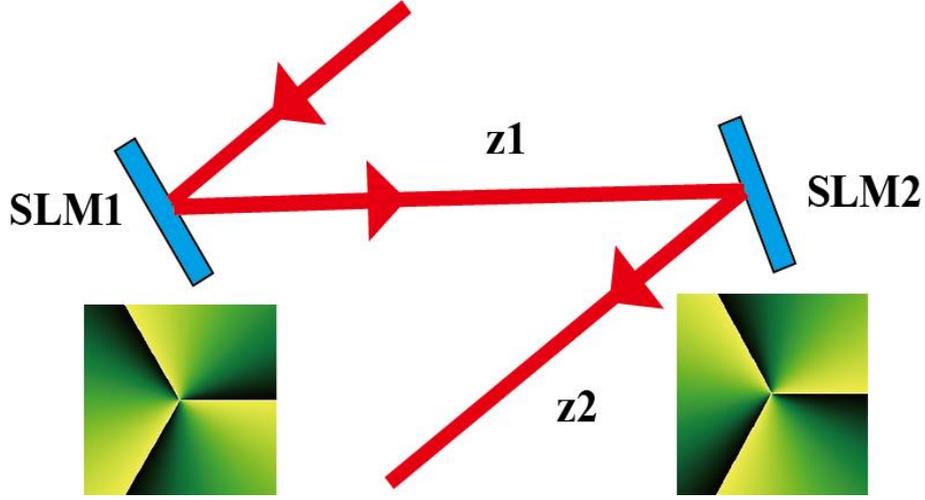

Fig.1. Simple diagram for vortex beam generation and transformation. SLM1, 2: spatial light modulator.

## 1. Generation and evolution of vortex beam from phase-only SLM

The practical fundamental beam for generating vortex beam is Gaussian beam, the amplitude of the Gaussian beam at the beam waist can be expressed as

$$E_0(r_0, \theta_0) = \sqrt{\frac{2}{\pi}} \exp(\frac{r^2}{w_0^2}) \qquad (1)$$

By applying a phase mask on the SLM of the form, the amplitude at the plane of SLM will become

$$E_1(r_0, \theta_0) = \sqrt{\frac{2}{\pi}} \exp(-\frac{r^2}{w_0^2}) \exp(-il\theta_0) \qquad (2)$$

After propagating away from the SLM, the beam amplitude can be obtained from Collins diffraction integral equation [9, 20]

$$E_2(r_1, \theta_1, z) = \frac{i}{\lambda B} \exp(-ikz_1) \int_0^{2\pi} \int_0^{\infty} E_1(r_0, \theta_0)$$
$$\exp\left\{-\frac{ik}{2B} \times [Ar_0^2 - 2rr_0 \cos(\theta_1 - \theta_0) + Dr^2]\right\} r_0 dr_0 d\theta_0 \qquad (3)$$

Where $z_1$ is the propagation distance, the ABCD matrix for free space propagation is

$$\begin{pmatrix} A & B \\ C & D \end{pmatrix} = \begin{pmatrix} 1 & z_1 \\ 0 & 1 \end{pmatrix} \qquad (4)$$

By inserting equation (2) (4) into (3), we obtain

$$E_2(r_1,\theta_1,z_1) = \frac{i^{l+1}}{2\lambda z_1}\sqrt{\frac{2}{\pi}}\frac{1}{w_0}\exp(-ikz_1)\exp(-il\theta_1)$$
$$\exp(-\frac{ik}{2z_1}r_1^2)\frac{b_1^l}{\varepsilon_1^{1+l/2}}F(l/2,l+1,\frac{b_1^2}{\varepsilon_1}) \qquad (5)$$

Equation (5) represents hyper geometric Gaussian mode. $F(\alpha,\beta,z)$ is the hyper geometric function. Where $b_1$ and $\varepsilon_1$ are defined as

$$b_1 = \frac{kr_1}{2z_1}, \varepsilon_1 = \frac{1}{w_0^2}+\frac{ik}{2z_1} \qquad (6)$$

## 2. Reversely transform vortex beam to Gaussian mode from phase-only SLM

After generation and evolution of the vortex beam, one usually need to reverse transform the beam to Gaussian mode. Below we will show how a vortex beam is reversely transformed by phase-only SLM and how the transformed beam gradually evolution to fundamental Gaussian mode. We first give the results for how a pure Lagurre-Gaussian (LG) mode is transformed by SLM. For a pure LG mode, when a reverse phase is applying on the SLM, the amplitude just on the SLM can be expressed as

$$E_2(r_1,\theta_1) = \sqrt{\frac{2}{\pi l!}}\frac{1}{w_0}(\frac{\sqrt{2}r_1}{w_0})^l \exp(-il\theta_1)\exp(il\theta_1) \qquad (7)$$

After propagating distance of $z_2$, by using equation (3), we can obtain the amplitude upon propagation.

$$E_3(r_2,\theta_2) = \frac{ik}{2z_2}\sqrt{\frac{2}{\pi}}\frac{1}{w_0}\exp(-ikz_2)\exp(-\frac{ik}{2z_2}r_2^2)\frac{b_2^l}{\varepsilon_2^{1+l/2}}F(-l/2,1,\frac{b_2^2}{\varepsilon_2}) \qquad (8)$$

Where definitions of $b_2$ and $\varepsilon_2$ are as following:

$$b_2 = \frac{kr_2}{2z_2}, \varepsilon_1 = \frac{1}{w_0^2}+\frac{ik}{2z_2} \qquad (9)$$

Equation (8) is accurate when the generated vortex has been propagating to the far field, then it can be viewed as pure LG mode.

To go one step further, if we considering the whole transformation processes, then we can derive the accurate field amplitude from Equation (5), after complex calculation, we obtain

$$E_3(r_2,\theta_2,z_1,z_2) = \frac{k^2}{2z_1z_2}\exp[-ik(z_1+z_2)]\exp(-\frac{ik}{2z_2}r_2^2)\frac{b_1'^l}{a_1^{1+l/2}}g^{-l/2}$$
$$\sum_{n=0}^{\infty}\frac{1}{n!}\frac{\Gamma(n+l/2)}{\Gamma(l/2)}\frac{\Gamma(l+1)}{\Gamma(n+l+1)}\Gamma(n+l/2+1)\left(\frac{\tau}{g}\right)^n F(n+l/2+1,1,-\frac{b_2^2}{g}) \qquad (10)$$

Where the definitions of parameters in the equation (10) are as following:

$$b_1' = \frac{k}{2z_1}, a_1 = \frac{1}{w_0^2}+\frac{ik}{2z_1}, \quad b_2 = \frac{kr_2}{2z_2}, g = \frac{b_1'^2}{a_1}+\frac{ik}{2z_1}+\frac{ik}{2z_2}, \tau = \frac{b_1'^2}{a_1} \qquad (11)$$

This concludes our theoretical calculations of the transformation of vortex beam generation,

evolution and reverse transformation of the beam to Gaussian mode.

## 3. Beam intensity distribution upon propagating

Based on the calculation above, we now give the transverse beam intensity upon propagating. For vortex beam generation and evolution, equation (5) is used in the calculations. In figure 2, the images shows the intensity distributions of the beam as function of propagating distance $z_1$. $l=9$, $w0=1.2$ mm are used in the calculations. We can see that the donut-shape gradually enlarges upon propagation, multi-rings appears in the outer region of the beam (see images in the left group of images). These rings gradually blur and disappear when propagating to the far field as show in image (i) of the left group of images. The right group images shows a zoomed view of the beam shape when the beam is propagating to a distance of 1-mm. We can conclude that pure LG mode is only an approximation of the beam generated in the far field.

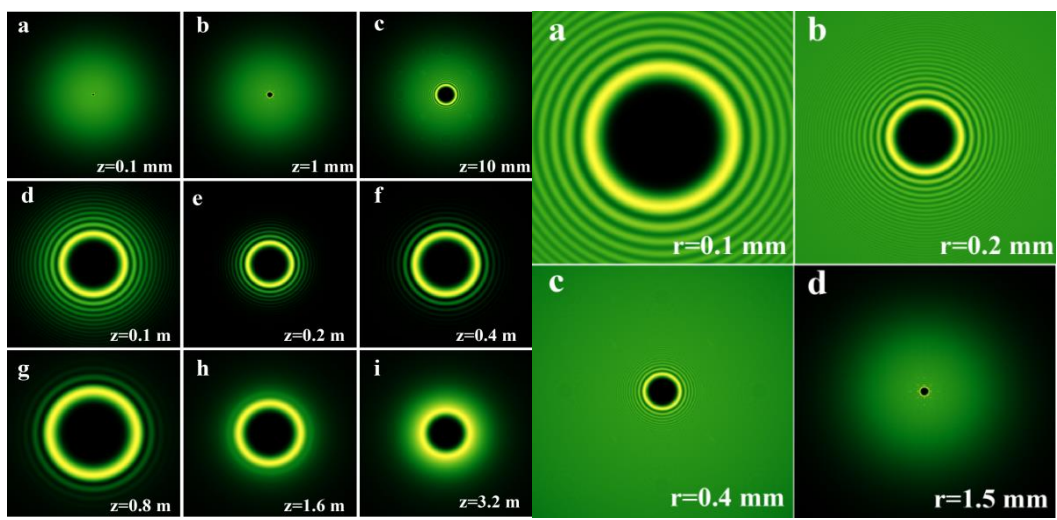

Fig. 2. Left group of images are the beam intensity distributions upon propagating distance for l=9, with initial beam waist of 1.2-mm. Right group of images are the zoomed views of image b in the left figure.

For superposition of vortex beam of 4 and -4, the propagation and evolution of the beam is showed in fig. 3. The evolution of superposition vortex beam shows the same properties as described for single vortex beam.

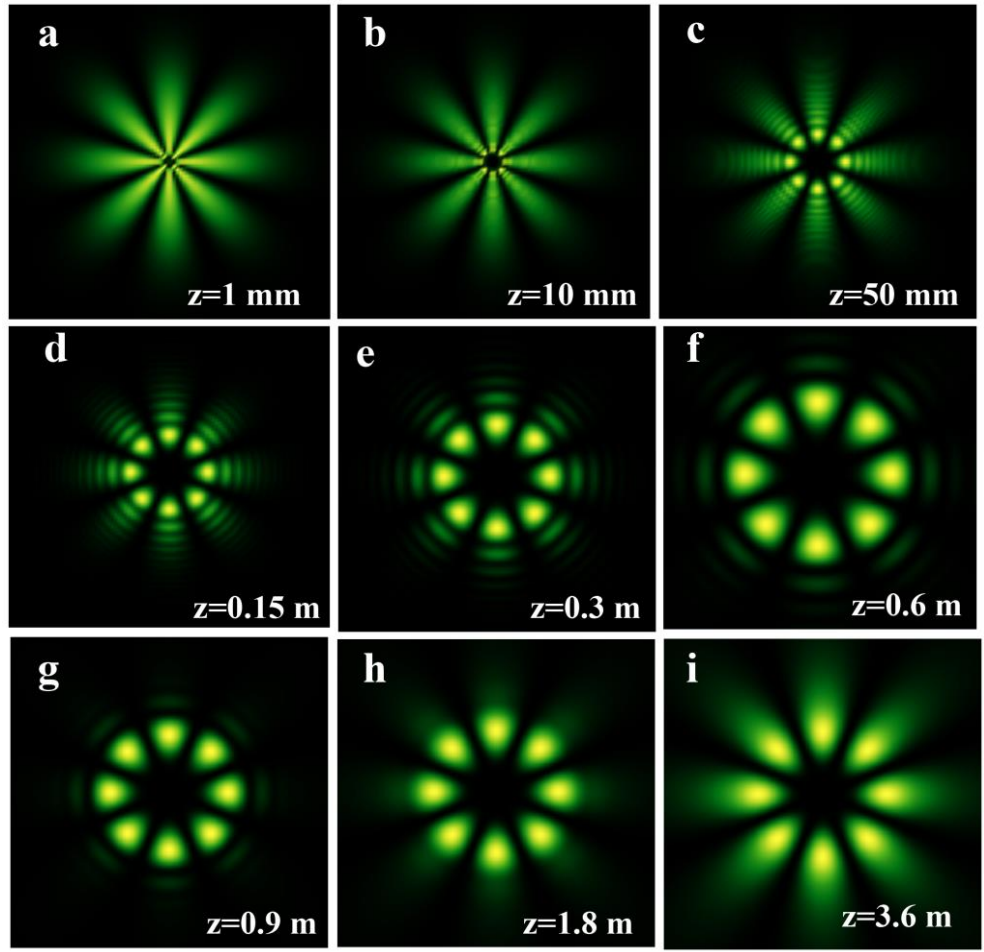

Fig. 3. Beam evolution behavior for superposition of vortex beam of 4 and -4, the initial beam waist is 1.0-mm.

We experimental verify the theoretical results by measuring the beam intensity distribution upon propagating distance for $l$=9 and superposition beam of 4 and -4, respectively. The beam waist used in the experiments is about 1.5-mm. The results are in well agreement with the theoretical results showed in Fig. 2 and 3. Because of limited optical table length and limited size of the window of the CCD camera, we cannot observe the intensity distributions in the far field regime, which makes image *i* in the right group of images slight different from the image showed in Fig. 3.

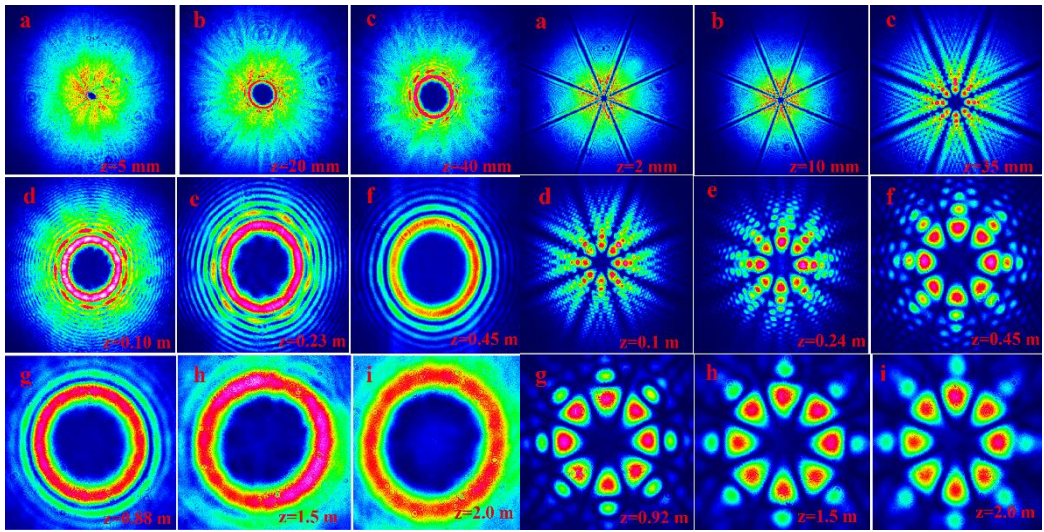

Fig. 4. The correspondingexperimental observed beam intensity distribution for Fig. 2 and 3 respectively. The fundamental Gaussian beam has a beam waist of 1.5-mm.

In many optical experiments, one need to reverse transform the vortex beam to Gaussian mode and coupling the beam to single mode fiber. Therefore, to precisely know the evolution behaviors of the beam in the reverse transformation is very important in experiment. The calculation is based on equation (8). The results are showed in figure 3, we can see that after applying a reverse phase in the SLM, the beam intensity distribution still has a ring shape upon small propagation distance, when the beam propagating away, a bright point start to emerge at the center of the beam, the intensity of the bright point become more and more stronger upon propagating. Finally, when the propagation distance is far enough, more than 95% power of the beam is focused at the central regime, which approximates a Gaussian distribution, this beam can be coupling to single mode fiber effectively.

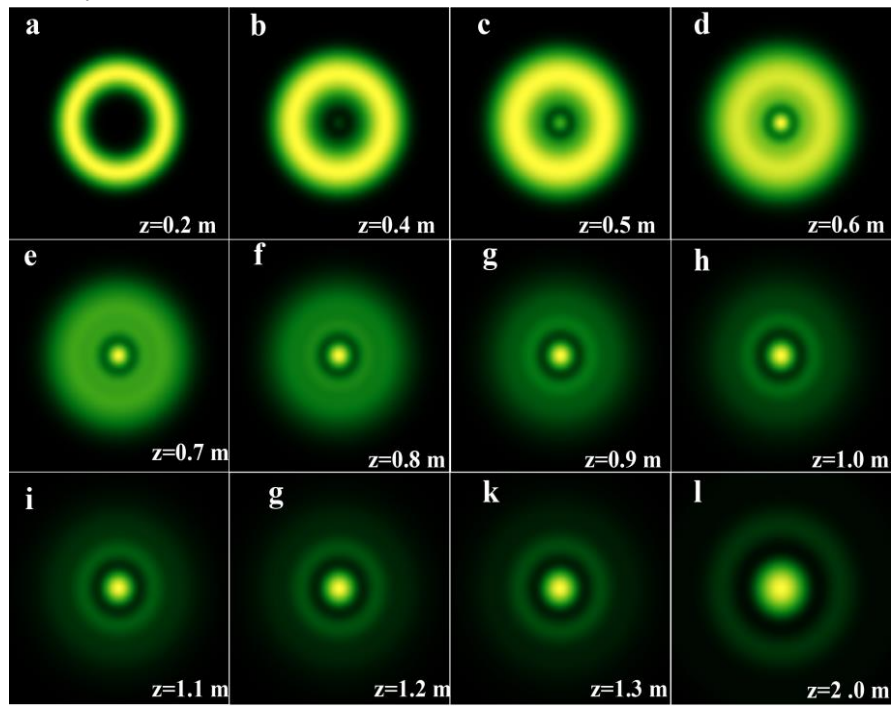

Fig. 5. Beam evolution in the reverse transformation of vortex beam l=9 based on equation (8). The beam waist used in the calculation is 0.4-mm.

## 4. Conclusion and outlook

We give a detail study of generation, evolution and reverse transformation of vortex beam by SLM based on diffraction theory. Analytical expressions are obtained for both cases which describe the whole beam evolution behaviors, we also experimentally measured the beam evolution using a CCD camera, these results are in well agreement with theoretical prediction. Analytical expressions obtained here can be extended to other experimental configurations including lenses in the transformation. Our results are very helpful for many optical experiments focused on optical vortex beam and OAM-mode multiplexed high capacity optical communications.

**Funding**

National Natural Science Foundation of China (Grant Nos. 11174271, 11604322, 61275115, 61435011, 61525504, 61605194) and the Fundamental Research Funds for the Central Universities.